\documentclass[10pt,a4paper,english]{article}\usepackage[]{graphicx}\usepackage[]{color}
\makeatletter
\def\maxwidth{ %
  \ifdim\Gin@nat@width>\linewidth
    \linewidth
  \else
    \Gin@nat@width
  \fi
}
\makeatother

\definecolor{fgcolor}{rgb}{0.345, 0.345, 0.345}
\newcommand{\hlnum}[1]{\textcolor[rgb]{0.686,0.059,0.569}{#1}}%
\newcommand{\hlstr}[1]{\textcolor[rgb]{0.192,0.494,0.8}{#1}}%
\newcommand{\hlcom}[1]{\textcolor[rgb]{0.678,0.584,0.686}{\textit{#1}}}%
\newcommand{\hlopt}[1]{\textcolor[rgb]{0,0,0}{#1}}%
\newcommand{\hlstd}[1]{\textcolor[rgb]{0.345,0.345,0.345}{#1}}%
\newcommand{\hlkwa}[1]{\textcolor[rgb]{0.161,0.373,0.58}{\textbf{#1}}}%
\newcommand{\hlkwb}[1]{\textcolor[rgb]{0.69,0.353,0.396}{#1}}%
\newcommand{\hlkwc}[1]{\textcolor[rgb]{0.333,0.667,0.333}{#1}}%
\newcommand{\hlkwd}[1]{\textcolor[rgb]{0.737,0.353,0.396}{\textbf{#1}}}%

\usepackage{framed}
\makeatletter
\newenvironment{kframe}{%
 \def\at@end@of@kframe{}%
 \ifinner\ifhmode%
  \def\at@end@of@kframe{\end{minipage}}%
  \begin{minipage}{\columnwidth}%
 \fi\fi%
 \def\FrameCommand##1{\hskip\@totalleftmargin \hskip-\fboxsep
 \colorbox{shadecolor}{##1}\hskip-\fboxsep
     \hskip-\linewidth \hskip-\@totalleftmargin \hskip\columnwidth}%
 \MakeFramed {\advance\hsize-\width
   \@totalleftmargin\z@ \linewidth\hsize
   \@setminipage}}%
 {\par\unskip\endMakeFramed%
 \at@end@of@kframe}
\makeatother

\definecolor{shadecolor}{rgb}{.97, .97, .97}
\definecolor{messagecolor}{rgb}{0, 0, 0}
\definecolor{warningcolor}{rgb}{1, 0, 1}
\definecolor{errorcolor}{rgb}{1, 0, 0}
\newenvironment{knitrout}{}{} % an empty environment to be redefined in TeX

\usepackage{alltt}

\usepackage[hyphens]{url}
\usepackage{amsmath}
\usepackage{amsfonts}
\usepackage{amssymb}
\usepackage{amsthm}

\theoremstyle{plain}

\theoremstyle{definition}

\theoremstyle{remark}

\PassOptionsToPackage{hyphens}{url}\usepackage{hyperref}
\usepackage{hyperref}
\usepackage[square,numbers]{natbib}
\usepackage[newitem,newenum,increaseonly]{paralist}


\usepackage{sadists}

    \usepackage{color} % Allow colors to be defined
    \usepackage{enumerate} % Needed for markdown enumerations to work
    \usepackage{amsmath} % Equations
    \usepackage{amssymb} % Equations
		\usepackage{fancyvrb} % verbatim replacement that allows latex
    \usepackage{hyperref}

    \definecolor{orange}{cmyk}{0,0.4,0.8,0.2}
    \definecolor{darkorange}{rgb}{.71,0.21,0.01}
    \definecolor{darkgreen}{rgb}{.12,.54,.11}
    \definecolor{myteal}{rgb}{.26, .44, .56}
    \definecolor{gray}{gray}{0.45}
    \definecolor{lightgray}{gray}{.95}
    \definecolor{mediumgray}{gray}{.8}
    \definecolor{inputbackground}{rgb}{.95, .95, .85}
    \definecolor{outputbackground}{rgb}{.95, .95, .95}
    \definecolor{traceback}{rgb}{1, .95, .95}
    \definecolor{red}{rgb}{.6,0,0}
    \definecolor{green}{rgb}{0,.65,0}
    \definecolor{brown}{rgb}{0.6,0.6,0}
    \definecolor{blue}{rgb}{0,.145,.698}
    \definecolor{purple}{rgb}{.698,.145,.698}
    \definecolor{cyan}{rgb}{0,.698,.698}
    \definecolor{lightgray}{gray}{0.5}
    
    \definecolor{darkgray}{gray}{0.25}
    \definecolor{lightred}{rgb}{1.0,0.39,0.28}
    \definecolor{lightgreen}{rgb}{0.48,0.99,0.0}
    \definecolor{lightblue}{rgb}{0.53,0.81,0.92}
    \definecolor{lightpurple}{rgb}{0.87,0.63,0.87}
    \definecolor{lightcyan}{rgb}{0.5,1.0,0.83}
    
    \DefineShortVerb[commandchars=\\\{\}]{\|}
    \DefineVerbatimEnvironment{Highlighting}{Verbatim}{commandchars=\\\{\}}
\makeatletter
\def\PY@reset{\let\PY@it=\relax \let\PY@bf=\relax%
    \let\PY@ul=\relax \let\PY@tc=\relax%
    \let\PY@bc=\relax \let\PY@ff=\relax}
\def\PY@tok#1{\csname PY@tok@#1\endcsname}
\def\PY@toks#1+{\ifx\relax#1\empty\else%
    \PY@tok{#1}\expandafter\PY@toks\fi}
\def\PY@do#1{\PY@bc{\PY@tc{\PY@ul{%
    \PY@it{\PY@bf{\PY@ff{#1}}}}}}}
\def\PY#1#2{\PY@reset\PY@toks#1+\relax+\PY@do{#2}}

\expandafter\def\csname PY@tok@gd\endcsname{\def\PY@tc##1{\textcolor[rgb]{0.63,0.00,0.00}{##1}}}
\expandafter\def\csname PY@tok@gu\endcsname{\let\PY@bf=\textbf\def\PY@tc##1{\textcolor[rgb]{0.50,0.00,0.50}{##1}}}
\expandafter\def\csname PY@tok@gt\endcsname{\def\PY@tc##1{\textcolor[rgb]{0.00,0.27,0.87}{##1}}}
\expandafter\def\csname PY@tok@gs\endcsname{\let\PY@bf=\textbf}
\expandafter\def\csname PY@tok@gr\endcsname{\def\PY@tc##1{\textcolor[rgb]{1.00,0.00,0.00}{##1}}}
\expandafter\def\csname PY@tok@cm\endcsname{\let\PY@it=\textit\def\PY@tc##1{\textcolor[rgb]{0.25,0.50,0.50}{##1}}}
\expandafter\def\csname PY@tok@vg\endcsname{\def\PY@tc##1{\textcolor[rgb]{0.10,0.09,0.49}{##1}}}
\expandafter\def\csname PY@tok@m\endcsname{\def\PY@tc##1{\textcolor[rgb]{0.40,0.40,0.40}{##1}}}
\expandafter\def\csname PY@tok@mh\endcsname{\def\PY@tc##1{\textcolor[rgb]{0.40,0.40,0.40}{##1}}}
\expandafter\def\csname PY@tok@go\endcsname{\def\PY@tc##1{\textcolor[rgb]{0.53,0.53,0.53}{##1}}}
\expandafter\def\csname PY@tok@ge\endcsname{\let\PY@it=\textit}
\expandafter\def\csname PY@tok@vc\endcsname{\def\PY@tc##1{\textcolor[rgb]{0.10,0.09,0.49}{##1}}}
\expandafter\def\csname PY@tok@il\endcsname{\def\PY@tc##1{\textcolor[rgb]{0.40,0.40,0.40}{##1}}}
\expandafter\def\csname PY@tok@cs\endcsname{\let\PY@it=\textit\def\PY@tc##1{\textcolor[rgb]{0.25,0.50,0.50}{##1}}}
\expandafter\def\csname PY@tok@cp\endcsname{\def\PY@tc##1{\textcolor[rgb]{0.74,0.48,0.00}{##1}}}
\expandafter\def\csname PY@tok@gi\endcsname{\def\PY@tc##1{\textcolor[rgb]{0.00,0.63,0.00}{##1}}}
\expandafter\def\csname PY@tok@gh\endcsname{\let\PY@bf=\textbf\def\PY@tc##1{\textcolor[rgb]{0.00,0.00,0.50}{##1}}}
\expandafter\def\csname PY@tok@ni\endcsname{\let\PY@bf=\textbf\def\PY@tc##1{\textcolor[rgb]{0.60,0.60,0.60}{##1}}}
\expandafter\def\csname PY@tok@nl\endcsname{\def\PY@tc##1{\textcolor[rgb]{0.63,0.63,0.00}{##1}}}
\expandafter\def\csname PY@tok@nn\endcsname{\let\PY@bf=\textbf\def\PY@tc##1{\textcolor[rgb]{0.00,0.00,1.00}{##1}}}
\expandafter\def\csname PY@tok@no\endcsname{\def\PY@tc##1{\textcolor[rgb]{0.53,0.00,0.00}{##1}}}
\expandafter\def\csname PY@tok@na\endcsname{\def\PY@tc##1{\textcolor[rgb]{0.49,0.56,0.16}{##1}}}
\expandafter\def\csname PY@tok@nb\endcsname{\def\PY@tc##1{\textcolor[rgb]{0.00,0.50,0.00}{##1}}}
\expandafter\def\csname PY@tok@nc\endcsname{\let\PY@bf=\textbf\def\PY@tc##1{\textcolor[rgb]{0.00,0.00,1.00}{##1}}}
\expandafter\def\csname PY@tok@nd\endcsname{\def\PY@tc##1{\textcolor[rgb]{0.67,0.13,1.00}{##1}}}
\expandafter\def\csname PY@tok@ne\endcsname{\let\PY@bf=\textbf\def\PY@tc##1{\textcolor[rgb]{0.82,0.25,0.23}{##1}}}
\expandafter\def\csname PY@tok@nf\endcsname{\def\PY@tc##1{\textcolor[rgb]{0.00,0.00,1.00}{##1}}}
\expandafter\def\csname PY@tok@si\endcsname{\let\PY@bf=\textbf\def\PY@tc##1{\textcolor[rgb]{0.73,0.40,0.53}{##1}}}
\expandafter\def\csname PY@tok@s2\endcsname{\def\PY@tc##1{\textcolor[rgb]{0.73,0.13,0.13}{##1}}}
\expandafter\def\csname PY@tok@vi\endcsname{\def\PY@tc##1{\textcolor[rgb]{0.10,0.09,0.49}{##1}}}
\expandafter\def\csname PY@tok@nt\endcsname{\let\PY@bf=\textbf\def\PY@tc##1{\textcolor[rgb]{0.00,0.50,0.00}{##1}}}
\expandafter\def\csname PY@tok@nv\endcsname{\def\PY@tc##1{\textcolor[rgb]{0.10,0.09,0.49}{##1}}}
\expandafter\def\csname PY@tok@s1\endcsname{\def\PY@tc##1{\textcolor[rgb]{0.73,0.13,0.13}{##1}}}
\expandafter\def\csname PY@tok@sh\endcsname{\def\PY@tc##1{\textcolor[rgb]{0.73,0.13,0.13}{##1}}}
\expandafter\def\csname PY@tok@sc\endcsname{\def\PY@tc##1{\textcolor[rgb]{0.73,0.13,0.13}{##1}}}
\expandafter\def\csname PY@tok@sx\endcsname{\def\PY@tc##1{\textcolor[rgb]{0.00,0.50,0.00}{##1}}}
\expandafter\def\csname PY@tok@bp\endcsname{\def\PY@tc##1{\textcolor[rgb]{0.00,0.50,0.00}{##1}}}
\expandafter\def\csname PY@tok@c1\endcsname{\let\PY@it=\textit\def\PY@tc##1{\textcolor[rgb]{0.25,0.50,0.50}{##1}}}
\expandafter\def\csname PY@tok@kc\endcsname{\let\PY@bf=\textbf\def\PY@tc##1{\textcolor[rgb]{0.00,0.50,0.00}{##1}}}
\expandafter\def\csname PY@tok@c\endcsname{\let\PY@it=\textit\def\PY@tc##1{\textcolor[rgb]{0.25,0.50,0.50}{##1}}}
\expandafter\def\csname PY@tok@mf\endcsname{\def\PY@tc##1{\textcolor[rgb]{0.40,0.40,0.40}{##1}}}
\expandafter\def\csname PY@tok@err\endcsname{\def\PY@bc##1{\setlength{\fboxsep}{0pt}\fcolorbox[rgb]{1.00,0.00,0.00}{1,1,1}{\strut ##1}}}
\expandafter\def\csname PY@tok@kd\endcsname{\let\PY@bf=\textbf\def\PY@tc##1{\textcolor[rgb]{0.00,0.50,0.00}{##1}}}
\expandafter\def\csname PY@tok@ss\endcsname{\def\PY@tc##1{\textcolor[rgb]{0.10,0.09,0.49}{##1}}}
\expandafter\def\csname PY@tok@sr\endcsname{\def\PY@tc##1{\textcolor[rgb]{0.73,0.40,0.53}{##1}}}
\expandafter\def\csname PY@tok@mo\endcsname{\def\PY@tc##1{\textcolor[rgb]{0.40,0.40,0.40}{##1}}}
\expandafter\def\csname PY@tok@kn\endcsname{\let\PY@bf=\textbf\def\PY@tc##1{\textcolor[rgb]{0.00,0.50,0.00}{##1}}}
\expandafter\def\csname PY@tok@mi\endcsname{\def\PY@tc##1{\textcolor[rgb]{0.40,0.40,0.40}{##1}}}
\expandafter\def\csname PY@tok@gp\endcsname{\let\PY@bf=\textbf\def\PY@tc##1{\textcolor[rgb]{0.00,0.00,0.50}{##1}}}
\expandafter\def\csname PY@tok@o\endcsname{\def\PY@tc##1{\textcolor[rgb]{0.40,0.40,0.40}{##1}}}
\expandafter\def\csname PY@tok@kr\endcsname{\let\PY@bf=\textbf\def\PY@tc##1{\textcolor[rgb]{0.00,0.50,0.00}{##1}}}
\expandafter\def\csname PY@tok@s\endcsname{\def\PY@tc##1{\textcolor[rgb]{0.73,0.13,0.13}{##1}}}
\expandafter\def\csname PY@tok@kp\endcsname{\def\PY@tc##1{\textcolor[rgb]{0.00,0.50,0.00}{##1}}}
\expandafter\def\csname PY@tok@w\endcsname{\def\PY@tc##1{\textcolor[rgb]{0.73,0.73,0.73}{##1}}}
\expandafter\def\csname PY@tok@kt\endcsname{\def\PY@tc##1{\textcolor[rgb]{0.69,0.00,0.25}{##1}}}
\expandafter\def\csname PY@tok@ow\endcsname{\let\PY@bf=\textbf\def\PY@tc##1{\textcolor[rgb]{0.67,0.13,1.00}{##1}}}
\expandafter\def\csname PY@tok@sb\endcsname{\def\PY@tc##1{\textcolor[rgb]{0.73,0.13,0.13}{##1}}}
\expandafter\def\csname PY@tok@k\endcsname{\let\PY@bf=\textbf\def\PY@tc##1{\textcolor[rgb]{0.00,0.50,0.00}{##1}}}
\expandafter\def\csname PY@tok@se\endcsname{\let\PY@bf=\textbf\def\PY@tc##1{\textcolor[rgb]{0.73,0.40,0.13}{##1}}}
\expandafter\def\csname PY@tok@sd\endcsname{\let\PY@it=\textit\def\PY@tc##1{\textcolor[rgb]{0.73,0.13,0.13}{##1}}}

\makeatother

    \definecolor{incolor}{rgb}{0.0, 0.0, 0.5}
    \definecolor{outcolor}{rgb}{0.545, 0.0, 0.0}

    \hypersetup{
      breaklinks=true,  % so long urls are correctly broken across lines
      colorlinks=true,
      urlcolor=blue,
      linkcolor=darkorange,
      citecolor=darkgreen,
      }
\IfFileExists{upquote.sty}{\usepackage{upquote}}{}
\begin{document}

\title{Moments of the log non-central chi-square distribution}
\author{Steven E. Pav \thanks{\email{spav@alumni.cmu.edu}}}
%\date{\today, \currenttime}

\maketitle
%UNFOLD

\providecommand{\df}[1][{}]{\mathUL{\nu}{}{#1}}
\providecommand{\Kgf}[1]{\MATHIT{K\wrapNeParens{#1}}}
\providecommand{\psif}[1]{\MATHIT{\psi\wrapNeParens{#1}}}
\providecommand{\polygam}[2][{1}]{\mathUL{\psi}{(#1)}{}\wrapNeParens{#2}}
\providecommand{\cumul}[1]{\mathUL{\kappa}{}{#1}}
\providecommand{\cmom}[1]{\mathUL{\mu'}{}{#1}}

%%%%%%%%%%%%%%%%%%%%%%%%%%%%%%%%%%%%%%%%%%%%%%%%%%%%%%%%%%%%%%%%%%%%%%%%
\begin{abstract}%FOLDUP
The cumulants and moments of the log of the non-central chi-square
distribution are derived. For example, the expected log of a 
chi-square random variable with \df degrees of freedom is
$\log{2} + \psif{\half[\df]}$. 
Applications to modeling probability distributions are discussed.
\end{abstract}%UNFOLD
\section{Introduction}%FOLDUP

The Edgeworth and Cornish-Fisher expansions allow one to approximate
the density, distribution, and quantile functions of probability
distributions whose cumulants are known. It is often remarked 
that these expansions are inaccurate for one-sided and highly skewed
probability distributions. \cite{1998A&AS..130..193B,edgeworthCF}

For example, consider a random variable which is the product of 
several chi-square random variables. The \kth{k} raw moment of
a chi-square random variable with \df degrees of freedom is
$2^k \fracc{\GAM{k + \df/2}}{\GAM{\df/2}}$. The raw moments of the
product of multiple independent chi-squares is simply the product
of the moments. The moments can then be converted to the cumulants.
The cumulants are then used in the Edgeworth expansion to approximate
the density. For even a small number of factors, the Edgeworth expansion
is inaccurate, sometimes yielding negative estimates, as illustrated 
in \figref{naiveprodchisq}. 

\begin{knitrout}\small
\definecolor{shadecolor}{rgb}{0.969, 0.969, 0.969}\color{fgcolor}\begin{kframe}
\begin{alltt}
\hlcom{# moments of chi-square}
\hlstd{chisq.moments} \hlkwb{<-} \hlkwa{function}\hlstd{(}\hlkwc{df}\hlstd{,} \hlkwc{order.max} \hlstd{=} \hlnum{3}\hlstd{) \{}
    \hlstd{orders} \hlkwb{<-} \hlnum{1}\hlopt{:}\hlstd{order.max}
    \hlstd{mu} \hlkwb{<-} \hlkwd{exp}\hlstd{(orders} \hlopt{*} \hlkwd{log}\hlstd{(}\hlnum{2}\hlstd{)} \hlopt{+} \hlkwd{lgamma}\hlstd{(orders} \hlopt{+}
        \hlstd{(df}\hlopt{/}\hlnum{2}\hlstd{))} \hlopt{-} \hlkwd{lgamma}\hlstd{(df}\hlopt{/}\hlnum{2}\hlstd{))}
    \hlkwd{return}\hlstd{(mu)}
\hlstd{\}}
\hlcom{# moments of product of chi-squares}
\hlstd{prodchisq.moments} \hlkwb{<-} \hlkwa{function}\hlstd{(}\hlkwc{dfs}\hlstd{,} \hlkwc{order.max} \hlstd{=} \hlnum{3}\hlstd{) \{}
    \hlstd{mu} \hlkwb{<-} \hlkwd{Reduce}\hlstd{(}\hlstr{"*"}\hlstd{,} \hlkwd{sapply}\hlstd{(dfs, chisq.moments,}
        \hlkwc{order.max} \hlstd{= order.max,} \hlkwc{simplify} \hlstd{=} \hlnum{FALSE}\hlstd{))}
    \hlkwd{return}\hlstd{(mu)}
\hlstd{\}}
\hlstd{rprodchisq} \hlkwb{<-} \hlkwa{function}\hlstd{(}\hlkwc{n}\hlstd{,} \hlkwc{dfs}\hlstd{) \{}
    \hlstd{X} \hlkwb{<-} \hlkwd{Reduce}\hlstd{(}\hlstr{"*"}\hlstd{,} \hlkwd{sapply}\hlstd{(dfs,} \hlkwa{function}\hlstd{(}\hlkwc{nu}\hlstd{) \{}
        \hlkwd{rchisq}\hlstd{(n,} \hlkwc{df} \hlstd{= nu)}
    \hlstd{\},} \hlkwc{simplify} \hlstd{=} \hlnum{FALSE}\hlstd{))}
    \hlkwd{return}\hlstd{(X)}
\hlstd{\}}
\hlkwd{require}\hlstd{(PDQutils)}
\hlstd{dprodchisq} \hlkwb{<-} \hlkwa{function}\hlstd{(}\hlkwc{x}\hlstd{,} \hlkwc{dfs}\hlstd{,} \hlkwc{log} \hlstd{=} \hlnum{FALSE}\hlstd{) \{}
    \hlstd{kappa} \hlkwb{<-} \hlstd{PDQutils}\hlopt{::}\hlkwd{moment2cumulant}\hlstd{(}\hlkwd{prodchisq.moments}\hlstd{(dfs,}
        \hlkwc{order.max} \hlstd{=} \hlnum{4}\hlstd{))}
    \hlstd{pdf} \hlkwb{<-} \hlkwd{dapx_edgeworth}\hlstd{(x, kappa,} \hlkwc{support} \hlstd{=} \hlkwd{c}\hlstd{(}\hlnum{0}\hlstd{,}
        \hlnum{Inf}\hlstd{),} \hlkwc{log} \hlstd{= log)}
    \hlkwd{return}\hlstd{(pdf)}
\hlstd{\}}

\hlkwd{require}\hlstd{(ggplot2)}
\hlstd{test_dens} \hlkwb{<-} \hlkwa{function}\hlstd{(}\hlkwc{dpqr}\hlstd{,} \hlkwc{nobs}\hlstd{,} \hlkwc{...}\hlstd{) \{}
    \hlstd{rv} \hlkwb{<-} \hlkwd{sort}\hlstd{(dpqr}\hlopt{$}\hlkwd{r}\hlstd{(nobs, ...))}
    \hlstd{data} \hlkwb{<-} \hlkwd{data.frame}\hlstd{(}\hlkwc{draws} \hlstd{= rv)}
    \hlstd{text.size} \hlkwb{<-} \hlnum{6}  \hlcom{# sigh}

    \hlcom{# http://stackoverflow.com/a/5688125/164611}
    \hlstd{p1} \hlkwb{<-} \hlkwd{qplot}\hlstd{(rv,} \hlkwc{geom} \hlstd{=} \hlstr{"blank"}\hlstd{)} \hlopt{+}
        \hlkwd{geom_line}\hlstd{(}\hlkwd{aes}\hlstd{(}\hlkwc{y} \hlstd{= ..density..,}
            \hlkwc{colour} \hlstd{=} \hlstr{"Empirical"}\hlstd{),} \hlkwc{stat} \hlstd{=} \hlstr{"density"}\hlstd{)} \hlopt{+}
        \hlkwd{stat_function}\hlstd{(}\hlkwc{fun} \hlstd{=} \hlkwa{function}\hlstd{(}\hlkwc{x}\hlstd{) \{}
            \hlstd{dpqr}\hlopt{$}\hlkwd{d}\hlstd{(x, ...)}
        \hlstd{\},} \hlkwd{aes}\hlstd{(}\hlkwc{colour} \hlstd{=} \hlstr{"Theoretical"}\hlstd{))} \hlopt{+}
        \hlkwd{geom_histogram}\hlstd{(}\hlkwd{aes}\hlstd{(}\hlkwc{y} \hlstd{= ..density..),}
            \hlkwc{alpha} \hlstd{=} \hlnum{0.3}\hlstd{)} \hlopt{+} \hlkwd{scale_colour_manual}\hlstd{(}\hlkwc{name} \hlstd{=} \hlstr{"Density"}\hlstd{,}
        \hlkwc{values} \hlstd{=} \hlkwd{c}\hlstd{(}\hlstr{"red"}\hlstd{,} \hlstr{"blue"}\hlstd{))} \hlopt{+}
        \hlkwd{theme}\hlstd{(}\hlkwc{text} \hlstd{=} \hlkwd{element_text}\hlstd{(}\hlkwc{size} \hlstd{= text.size))} \hlopt{+}
        \hlkwd{labs}\hlstd{(}\hlkwc{title} \hlstd{=} \hlstr{"Density (tests dfunc)"}\hlstd{)}
    \hlkwd{print}\hlstd{(p1)}
\hlstd{\}}

\hlstd{dfs} \hlkwb{<-} \hlkwd{c}\hlstd{(}\hlnum{40}\hlstd{,} \hlnum{30}\hlstd{,} \hlnum{50}\hlstd{,} \hlnum{20}\hlstd{,} \hlnum{10}\hlstd{)}
\hlkwd{test_dens}\hlstd{(}\hlkwd{list}\hlstd{(}\hlkwc{r} \hlstd{= rprodchisq,} \hlkwc{d} \hlstd{= dprodchisq),}
    \hlkwc{nobs} \hlstd{=} \hlnum{2}\hlopt{^}\hlnum{14}\hlstd{, dfs)}
\end{alltt}
\end{kframe}\begin{figure}
\includegraphics[width=5in,height=3in]{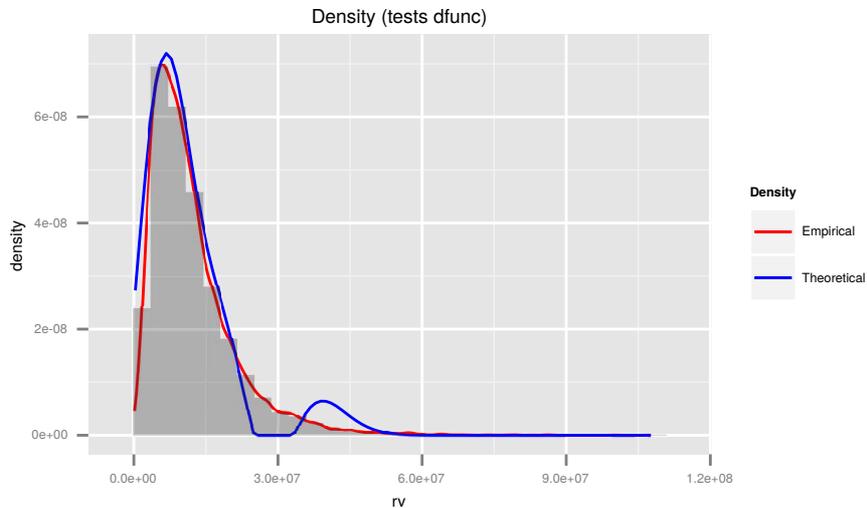} \caption[The Edgeworth expansion is inaccurate for the product of chi-squares variates]{The Edgeworth expansion is inaccurate for the product of chi-squares variates.}\label{fig:naiveprodchisq}
\end{figure}

\end{knitrout}

% 2FIX

%UNFOLD
\section{Moments of the log chi-square distribution}%FOLDUP

One possible fix to this problem is to use the Edgeworth
expansions to approximate the density of the \emph{log}
of the chi-square, or the weighted sum of logs of chi-square
variates.  This latter approach would allow one to model
the doubly non-central F distribution, for example, as well
as products of arbitrary chi-squares to different powers.

Let $x\sim\chisqlaw{\df}$ be a chi-square variate with \df
degrees of freedom. Let $y=\log{x}$. Consider the cumulant
generating function of $y$. It is 	
$$
\Kgf{t} = \log{\E{\exp{yt}}} = \log{\E{x^t}}.
$$
The moments of the central chi-square are known \cite{walck:1996},
yielding the expression
$$
\Kgf{t} = t \log{2} + \log{\GAM{\half[\df] + t}} - \log{\GAM{\half[\df]}}.
$$
The \kth{k} cumulant of $y$ is the \kth{k} derivative evaluated at $t=0$.
The derivative of the log of the Gamma function is the `psi' function,
$\psif{\cdot}$, while its derivatives are the `polygamma' functions,
$\polygam[n]{\cdot}$.  \cite[6.3, 6.4]{abramowitz_stegun}
Letting \cumul{j} be the \kth{j} raw cumulant of $y$, we have
\begin{equation}
\label{eqn:chisq_cumuls}
\cumul{j} = \begin{cases}
\log{2} + \psif{\half[\df]},& \text{if } j = 1,\\
\polygam[j-1]{\half[\df]},& \text{if } j > 1.
\end{cases}
\end{equation}

The moments can then be computed from the cumulants via the usual
formula:
\begin{equation}
\cmom{n} = \cumul{n} + \sum_{m=1}^{n-1} {n-1 \choose m-1} \cumul{m} \cmom{n-m}.
\end{equation}
Note that the first raw cumulant equals the first moment.  That is,
\begin{equation}
\E{y} = \log{2} + \psif{\half[\df]}.
\end{equation}

%UNFOLD
\section{Moments of the log non-central chi-square distribution}%FOLDUP

To compute the moments of the log of the \emph{non-central} chi-square,
the crucial observation is that the density of the non-central chi-square
can be expressed as a Poisson mixture of central 
chi-squares.  \cite{walck:1996} That is, 
if \chisqpdf{x}{\df} is the density of the central chi-square distribution
with \df degrees of freedom, and \nchisqpdf{x}{\nccp,\df} is the density
of the non-central chi-square with \df degrees of freedom and non-centrality
parameter \nccp, then 
\begin{equation}
\nchisqpdf{x}{\nccp,\df} = \sum_{j=0}^{\infty} \exp{-\halff[\nccp]}
\frac{\wrapParens{\half[\nccp]}^j}{j!} \chisqpdf{x}{\df + 2j}.
\end{equation}

Using the change of variables formula, if $y$ is the log of a non-central
chi-square variate with \df degrees of freedom and non-centrality parameter
\nccp, then the density of $y$ follows a similar relationship:
\begin{equation}
\FOOpdf{Y}{y}{\nccp,\df} = 
\exp{y} \nchisqpdf{\exp{y}}{\nccp,\df} = 
\exp{y} \sum_{j=0}^{\infty} \exp{-\halff[\nccp]}
\frac{\wrapParens{\half[\nccp]}^j}{j!} \chisqpdf{\exp{y}}{\df + 2j}.
\end{equation}

Because the uncentered moments are defined in terms of an integral, which
is a linear operator, the moments of $y$ can be expressed as a similar sum:
\begin{equation}
\begin{split}
\E{y^k} &= 
\sum_{j=0}^{\infty} \exp{-\halff[\nccp]} 
\frac{\wrapParens{\half[\nccp]}^j}{j!} 
\int_{-\infty}^{\infty} \exp{z} \chisqpdf{\exp{z}}{\df + 2j} z^k \dx[z],\\
&=
\sum_{j=0}^{\infty} \exp{-\halff[\nccp]} 
\frac{\wrapParens{\half[\nccp]}^j}{j!} \cmom{k,\df+2j},
\end{split}
\end{equation}
where \cmom{k,\df+2j} is the \kth{k} moment of the log of the central
chi-square distribution with $\df+2j$ degrees of freedom.

For the particular case of $k=1$, because the first cumulant is the first
moment, via \eqnref{chisq_cumuls}, we have
\begin{equation}
\E{y} = \log{2} + \sum_{j=0}^{\infty} \exp{-\halff[\nccp]} 
\frac{\wrapParens{\half[\nccp]}^j}{j!} \psif{j + \halff[\df]}.
\end{equation}

Note that this does not seem to match the equations given by 
Moser, even in the simple case $\nccp=0$. \cite{moser2004}
(It appears that Moser's result is missing a summand of $\log{2}$.)
It is easy to check this formula via Monte Carlo simulations,
as below. The results of these experiments, reported in 
\tabref{mc_check} indicate that the equations are
indeed accurate.

\begin{kframe}
\begin{alltt}
\hlkwd{require}\hlstd{(PDQutils)}
\hlcom{# cumulants of the log of the central}
\hlcom{# chi-square}
\hlstd{lc_cumuls} \hlkwb{<-} \hlkwa{function}\hlstd{(}\hlkwc{df}\hlstd{,} \hlkwc{order.max} \hlstd{=} \hlnum{3}\hlstd{,}
    \hlkwc{orders} \hlstd{=} \hlkwd{c}\hlstd{(}\hlnum{1}\hlopt{:}\hlstd{order.max)) \{}
    \hlstd{kappa} \hlkwb{<-} \hlkwd{psigamma}\hlstd{(df}\hlopt{/}\hlnum{2}\hlstd{,} \hlkwc{deriv} \hlstd{= orders} \hlopt{-}
        \hlnum{1}\hlstd{)}
    \hlstd{kappa[}\hlnum{1}\hlstd{]} \hlkwb{<-} \hlstd{kappa[}\hlnum{1}\hlstd{]} \hlopt{+} \hlkwd{log}\hlstd{(}\hlnum{2}\hlstd{)}
    \hlkwd{return}\hlstd{(kappa)}
\hlstd{\}}
\hlcom{# moments of the log of the central}
\hlcom{# chi-square}
\hlstd{lc_moments} \hlkwb{<-} \hlkwa{function}\hlstd{(}\hlkwc{df}\hlstd{,} \hlkwc{order.max} \hlstd{=} \hlnum{3}\hlstd{,}
    \hlkwc{orders} \hlstd{=} \hlkwd{c}\hlstd{(}\hlnum{1}\hlopt{:}\hlstd{order.max)) \{}
    \hlstd{kappa} \hlkwb{<-} \hlkwd{lc_cumuls}\hlstd{(df,} \hlkwc{orders} \hlstd{= orders)}
    \hlstd{mu} \hlkwb{<-} \hlstd{PDQutils}\hlopt{::}\hlkwd{cumulant2moment}\hlstd{(kappa)}
    \hlkwd{return}\hlstd{(mu)}
\hlstd{\}}
\hlcom{# moments of the log of the}
\hlcom{# non-central chi-square}
\hlstd{lnc_moments} \hlkwb{<-} \hlkwa{function}\hlstd{(}\hlkwc{df}\hlstd{,} \hlkwc{ncp} \hlstd{=} \hlnum{0}\hlstd{,}
    \hlkwc{order.max} \hlstd{=} \hlnum{3}\hlstd{,} \hlkwc{orders} \hlstd{=} \hlkwd{c}\hlstd{(}\hlnum{1}\hlopt{:}\hlstd{order.max)) \{}
    \hlkwd{stopifnot}\hlstd{(ncp} \hlopt{>=} \hlnum{0}\hlstd{)}
    \hlkwa{if} \hlstd{(ncp} \hlopt{>} \hlnum{0}\hlstd{) \{}
        \hlstd{hancp} \hlkwb{<-} \hlstd{ncp}\hlopt{/}\hlnum{2}
        \hlcom{# should be smarter about 0:100 here.}
        \hlstd{allmu} \hlkwb{<-} \hlkwd{sapply}\hlstd{(}\hlnum{0}\hlopt{:}\hlnum{100}\hlstd{,} \hlkwa{function}\hlstd{(}\hlkwc{iv}\hlstd{) \{}
            \hlkwd{exp}\hlstd{(}\hlopt{-}\hlstd{hancp} \hlopt{+} \hlstd{iv} \hlopt{*} \hlkwd{log}\hlstd{(hancp)} \hlopt{-}
                \hlkwd{lfactorial}\hlstd{(iv))} \hlopt{*} \hlkwd{lc_moments}\hlstd{(df} \hlopt{+}
                \hlnum{2} \hlopt{*} \hlstd{iv,} \hlkwc{orders} \hlstd{= orders)}
        \hlstd{\},} \hlkwc{simplify} \hlstd{=} \hlnum{FALSE}\hlstd{)}
        \hlstd{mu} \hlkwb{<-} \hlkwd{Reduce}\hlstd{(}\hlstr{"+"}\hlstd{, allmu)}
    \hlstd{\}} \hlkwa{else} \hlstd{\{}
        \hlstd{mu} \hlkwb{<-} \hlkwd{lc_moments}\hlstd{(}\hlkwc{df} \hlstd{= df,} \hlkwc{orders} \hlstd{= orders)}
    \hlstd{\}}
    \hlkwd{return}\hlstd{(mu)}
\hlstd{\}}
\hlkwd{set.seed}\hlstd{(}\hlnum{1231591}\hlstd{)}
\hlstd{df} \hlkwb{<-} \hlnum{50}
\hlstd{ncp} \hlkwb{<-} \hlnum{1.5}
\hlstd{nsim} \hlkwb{<-} \hlnum{1e+06}
\hlstd{x} \hlkwb{<-} \hlkwd{rchisq}\hlstd{(nsim,} \hlkwc{df} \hlstd{= df,} \hlkwc{ncp} \hlstd{= ncp)}
\hlstd{y} \hlkwb{<-} \hlkwd{log}\hlstd{(x)}

\hlstd{nord} \hlkwb{<-} \hlnum{6}
\hlstd{empirical.mu} \hlkwb{<-} \hlkwd{sapply}\hlstd{(}\hlnum{1}\hlopt{:}\hlstd{nord,} \hlkwa{function}\hlstd{(}\hlkwc{k}\hlstd{) \{}
    \hlkwd{mean}\hlstd{(y}\hlopt{^}\hlstd{k)}
\hlstd{\})}
\hlstd{theoretical.mu} \hlkwb{<-} \hlkwd{lnc_moments}\hlstd{(}\hlkwc{df} \hlstd{= df,}
    \hlkwc{ncp} \hlstd{= ncp,} \hlkwc{order.max} \hlstd{= nord)}
\end{alltt}
\end{kframe}
% latex table generated in R 3.1.3 by xtable 1.7-4 package
% Fri Mar 20 22:08:04 2015
\begin{table}[ht]
\centering
\begin{tabular}{rrr}
  \hline
order & empirical & theoretical \\ 
  \hline
  1 & 3.92 & 3.92 \\ 
    2 & 15.42 & 15.42 \\ 
    3 & 60.79 & 60.78 \\ 
    4 & 240.26 & 240.22 \\ 
    5 & 951.97 & 951.78 \\ 
    6 & 3781.22 & 3780.36 \\ 
   \hline
\end{tabular}
\caption{Empirical and theoretical moments up to order 6 are shown for $10^6$ draws from the log of a non-central chi-square distribution with 50 degrees of freedom and non-centrality parameter 1.5.} 
\label{tab:mc_check}
\end{table}

%UNFOLD
\section{Using the Moments}

While the moments computation is perhaps of theoretical interest, the nominal
impetus for this work was more accurate simulation of the density of products
of non-central chi-squares taken to powers. Here we first approximate the
density of the log of such a distribution, using additivity of cumulants, 
via an Edgeworth expansion, then use change of variables to recover the density
of the product of chi-squares. The resultant density estimator, shown in 
\figref{betterprodchisq}, is an improvement, at least under an `eyeball test.'

\begin{knitrout}\small
\definecolor{shadecolor}{rgb}{0.969, 0.969, 0.969}\color{fgcolor}\begin{kframe}
\begin{alltt}
\hlcom{# cumulants of the log of the}
\hlcom{# non-central chi-square}
\hlstd{lnc_cumuls} \hlkwb{<-} \hlkwa{function}\hlstd{(}\hlkwc{df}\hlstd{,} \hlkwc{ncp} \hlstd{=} \hlnum{0}\hlstd{,} \hlkwc{order.max} \hlstd{=} \hlnum{3}\hlstd{,}
    \hlkwc{orders} \hlstd{=} \hlkwd{c}\hlstd{(}\hlnum{1}\hlopt{:}\hlstd{order.max)) \{}
    \hlstd{mu} \hlkwb{<-} \hlkwd{lnc_moments}\hlstd{(df, ncp,} \hlkwc{orders} \hlstd{= orders)}
    \hlstd{kappa} \hlkwb{<-} \hlstd{PDQutils}\hlopt{::}\hlkwd{moment2cumulant}\hlstd{(mu)}
    \hlkwd{return}\hlstd{(kappa)}
\hlstd{\}}
\hlcom{# compute the cumulants of the}
\hlcom{# sumlogchisq distribution.}
\hlstd{sumlogchisq_cumuls} \hlkwb{<-} \hlkwa{function}\hlstd{(}\hlkwc{wts}\hlstd{,} \hlkwc{df}\hlstd{,}
    \hlkwc{ncp} \hlstd{=} \hlnum{0}\hlstd{,} \hlkwc{order.max} \hlstd{=} \hlnum{3}\hlstd{) \{}
    \hlstd{subkappa} \hlkwb{<-} \hlkwd{mapply}\hlstd{(}\hlkwa{function}\hlstd{(}\hlkwc{w}\hlstd{,} \hlkwc{dd}\hlstd{,}
        \hlkwc{nn}\hlstd{) \{}
        \hlstd{(w}\hlopt{^}\hlstd{(}\hlnum{1}\hlopt{:}\hlstd{order.max))} \hlopt{*} \hlkwd{lnc_cumuls}\hlstd{(}\hlkwc{df} \hlstd{= dd,}
            \hlkwc{ncp} \hlstd{= nn,} \hlkwc{order.max} \hlstd{= order.max)}
    \hlstd{\}, wts, df, ncp,} \hlkwc{SIMPLIFY} \hlstd{=} \hlnum{FALSE}\hlstd{)}
    \hlstd{kappa} \hlkwb{<-} \hlkwd{Reduce}\hlstd{(}\hlstr{"+"}\hlstd{, subkappa)}
    \hlkwd{return}\hlstd{(kappa)}
\hlstd{\}}
\hlstd{dsumlogchisq} \hlkwb{<-} \hlkwa{function}\hlstd{(}\hlkwc{x}\hlstd{,} \hlkwc{wts}\hlstd{,} \hlkwc{df}\hlstd{,}
    \hlkwc{ncp} \hlstd{=} \hlnum{0}\hlstd{,} \hlkwc{log} \hlstd{=} \hlnum{FALSE}\hlstd{,} \hlkwc{order.max} \hlstd{=} \hlnum{6}\hlstd{) \{}
    \hlstd{kappa} \hlkwb{<-} \hlkwd{sumlogchisq_cumuls}\hlstd{(wts,}
        \hlstd{df, ncp,} \hlkwc{order.max} \hlstd{= order.max)}
    \hlstd{retval} \hlkwb{<-} \hlstd{PDQutils}\hlopt{::}\hlkwd{dapx_edgeworth}\hlstd{(x,}
        \hlstd{kappa,} \hlkwc{log} \hlstd{= log)}
    \hlkwd{return}\hlstd{(retval)}
\hlstd{\}}

\hlcom{# use change of variables:}
\hlstd{dprodchisq2} \hlkwb{<-} \hlkwa{function}\hlstd{(}\hlkwc{x}\hlstd{,} \hlkwc{dfs}\hlstd{,} \hlkwc{log} \hlstd{=} \hlnum{FALSE}\hlstd{) \{}
    \hlstd{dx} \hlkwb{<-} \hlkwd{dsumlogchisq}\hlstd{(}\hlkwd{log}\hlstd{(x),} \hlkwc{wts} \hlstd{=} \hlnum{1}\hlstd{,}
        \hlkwc{df} \hlstd{= dfs,} \hlkwc{ncp} \hlstd{=} \hlnum{0}\hlstd{,} \hlkwc{log} \hlstd{= log)}
    \hlkwa{if} \hlstd{(log) \{}
        \hlstd{dx} \hlkwb{<-} \hlstd{dx} \hlopt{-} \hlkwd{log}\hlstd{(x)}
    \hlstd{\}} \hlkwa{else} \hlstd{\{}
        \hlstd{dx} \hlkwb{<-} \hlstd{dx}\hlopt{/}\hlstd{x}
    \hlstd{\}}
    \hlkwd{return}\hlstd{(dx)}
\hlstd{\}}

\hlstd{dfs} \hlkwb{<-} \hlkwd{c}\hlstd{(}\hlnum{40}\hlstd{,} \hlnum{30}\hlstd{,} \hlnum{50}\hlstd{,} \hlnum{20}\hlstd{,} \hlnum{10}\hlstd{)}
\hlkwd{test_dens}\hlstd{(}\hlkwd{list}\hlstd{(}\hlkwc{r} \hlstd{= rprodchisq,} \hlkwc{d} \hlstd{= dprodchisq2),}
    \hlkwc{nobs} \hlstd{=} \hlnum{2}\hlopt{^}\hlnum{14}\hlstd{, dfs)}
\end{alltt}
\end{kframe}\begin{figure}
\includegraphics[width=5in,height=3in]{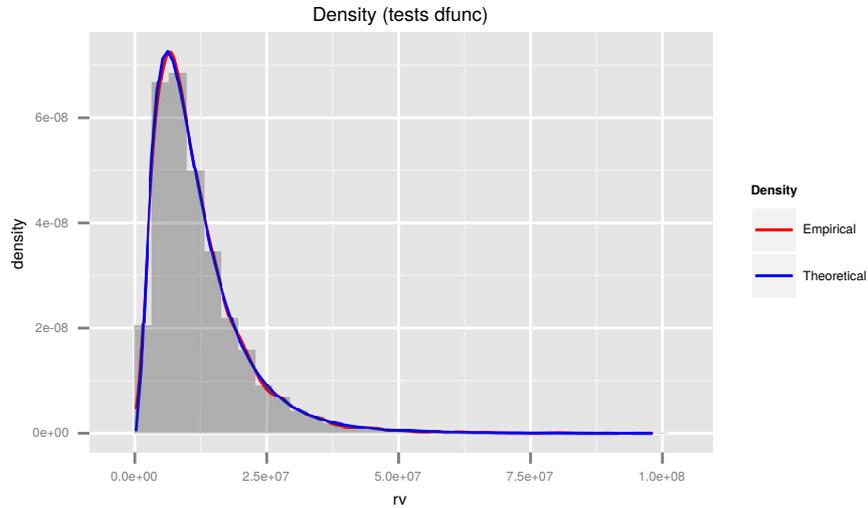} \caption[Applying the Edgeworth expansion to the sum of log chi-squares, then performing a change of variables, results in a more accurate density estimate]{Applying the Edgeworth expansion to the sum of log chi-squares, then performing a change of variables, results in a more accurate density estimate.}\label{fig:betterprodchisq}
\end{figure}

\end{knitrout}

\nocite{walck:1996,moser2004,PDQutils-Manual}

%%%%%%%%%%%%%%%%%%%%%%%%%%%%%%%%%%%%%%%%%%%%%%%%%%%%%%%%%%%%%%%%%%%%%%%%
% bibliography%FOLDUP
%\bibliographystyle{jss}
%\bibliographystyle{siam}
%\bibliographystyle{ieeetr}
\bibliographystyle{plainnat}
\bibliography{sadists,rauto}
%UNFOLD

\end{document}